\documentclass[onecolumn,showpacs,preprintnumbers,amsmath,amssymb]{revtex4}

\usepackage{graphicx}% Include figure files
\usepackage{dcolumn}% Align table columns on decimal point
\usepackage{bm}% bold math

\begin{document}

\preprint{APS/123-QED}

\title{Comparative Study of Different Discretizations of the $\phi^4$
Model}

\author{Ishani Roy$^{1,2}$, Sergey V. Dmitriev$^3$,
Panayotis G. Kevrekidis$^1$ and Avadh Saxena$^2$}

\affiliation{$^1$ Department of Mathematics and Statistics,
University of Massachusetts, Amherst, MA 01003-4515, USA \\
$^2$ Center for Nonlinear Studies and Theoretical Division, Los
Alamos National Laboratory, Los Alamos, New Mexico 87545, USA \\
 $^3$ Institute of Industrial Science, the University of
Tokyo, Komaba, Meguro-ku, Tokyo 153-8505, Japan }
\date{\today}

\begin{abstract}
We examine various recently proposed discretizations of the
well-known $\phi^4$ field theory. We compare and contrast the
properties of their fundamental solutions including the nature of
their kink-type solitary waves and the spectral properties of the
linearization around such waves. We study these features as a
function of the lattice spacing $h$, as one deviates from the
continuum limit of $h \rightarrow 0$. We then proceed to a more
``stringent'' comparison of the models, by discussing the
scattering properties of a kink-antikink pair for the different
discretizations. These collisions are well-known to possess
properties that quite sensitively depend on the initial speed even
at the continuum limit. We examine how typical model behaviors are
modified in the presence (and as a function) of discreteness and
attempt to extract qualitative trends and issue pertinent warnings
about some of the surprising resulting properties.
\end{abstract}


\maketitle

%--------------------------------------------------------------------------
\section{Introduction} \label{sec:Introduction}
%--------------------------------------------------------------------------

In the past few years, a variety of physical applications ranging
from Bose-Einstein condensates in optical lattices \cite{konotop},
to arrays of waveguides in nonlinear optics \cite{nlo} and even to
the dynamics of DNA \cite{peyrard} have stimulated an enormous
growth in the study of discrete models and the
differential-difference equations that describe them. These models
have a two-fold role and importance. On the one hand, they serve
as discretizations of the corresponding continuum field theories;
however, on the other hand, they may also be important physical
models in their own right, e.g. in the context of crystal lattices.

An important class of models (and discretizations thereof) that is
relevant to a wide variety of applications consists of the
so-called Klein-Gordon type equations \cite{kivshar1}. One of the
particularly interesting equations within this family is the
so-called  $\phi^4$ model \cite{belova}, featuring a wave equation
with a cubic nonlinear (odd-power) polynomial added to it. This
model has been physically argued as being of relevance in
describing cosmic domain walls in cosmological settings
\cite{anninos}, but also structural phase transitions, uniaxial
ferroelectrics or even simple polymeric chains; see e.g.
\cite{Campbell} and references therein. A particularly intriguing
feature that was discovered early on in the continuum limit was
the existence of a fractal structure \cite{anninos} in the
collisions between the fundamental nonlinear wave structures (a
kink and an anti-kink) in this model. This is a topic that was
initiated by the numerical investigations of Refs.
\cite{Campbell,Campbell1} (see also \cite{belova}) and is a topic
that is still under active investigation (see e.g., the very
recent mathematical analysis of the relevant mechanism provided in
Ref. \cite{goodman}).

On the other hand, more recently, an issue that has concerned
research work has been how to produce discretizations of such
continuum models (such as the $\phi^4$ model or its complex
cousin, the nonlinear-Schr{\"o}dinger equation \cite{sulem}) that
preserve some of the important properties of the corresponding
continuum limit. One of the non-trivial aspects of this endeavor
is the generation of discrete models in space that maintain some
of the key invariances of their continuum siblings. For instance,
in the uniform continuum medium, solutions can be shifted
arbitrarily along a certain direction $x$ by any $x_0$ ($x$ is the
spatial coordinate and $x_{0}={\rm const.}$), due to the
underlying translational invariance. However, discretizations
generically will fail to maintain that feature and in the most
straightforward versions thereof, equilibrium static solutions
exist for a discrete rather than a continuum set of $x_0$
\cite{kivshar1}. Some of these equilibrium solutions correspond to
energy maxima and are unstable, while others, corresponding to
energy minima, are stable. The difference between such maxima
and minima of the energy is typically referred to as the
Peierls-Nabarro barrier (PNb). One of the topics of intense
research efforts in the past few years has been to develop
discretizations that do not present such energetic barriers; this
is done in the hope that the latter class of models may provide more
faithful representations of their continuum counterparts,
regarding both symmetry properties and traveling solution
features.

The result of the above considerations in recent years has been
the systematic construction of a large class of non-integrable
discrete Klein-Gordon equations free of the Peierls-Nabarro
barrier (PNb-free). Since the standard discretization of the
continuum $\phi^4$ models conserves the energy but does have a
PNb, Speight and co-workers \cite{SpeightKleinGordon} (see also
\cite{Speight,SpeightPhi4}) originally used a Bogomol'nyi argument
\cite{Bogom}, in order to eliminate that barrier. A later
successful attempt (that produced multiple PNb-free models) was
based on a different perspective, namely the one of associating
the PNb-free models to momentum-conserving discretizations
\cite{KevrekidisPhysD}. Subsequently, yet another such model was
recently proposed in \cite{Saxena}. Furthermore, these approaches
were systematized and generalized through their formulation by
means of a two-point discrete version of the first integral of the
static continuum Klein-Gordon equation \cite{JPhysA,Barashenkov}.
We should note in passing that similar discretization efforts have
recently been extended to the nonlinear Schr{\"o}dinger equation
\cite{dk,dep,krss}.

One of the important questions that naturally emerges in the
presence of this extensive recent literature is, indeed, how
accurately do we expect these models to track the continuum limit
behavior and how various properties are affected by the
discreteness, as a function of its characteristic parameter (the
lattice spacing $h$). Already, to some extent, there have been
concerns regarding that question in that simulations of more
sensitive phenomena such as kink collisions in the ``Speight
discretization'' \cite{SpeightKleinGordon,SpeightPhi4} were only
faithful to the continuum limit for fine lattices ($h \approx 0.1$
or less) \cite{almeida}. Here, we examine this question a bit more
broadly and in more detail through numerical computations and
analytical considerations comparing/contrasting five different
discretizations of the continuum $\phi^4$ field theory. Among them
is the ``standard'', classical $\phi^4$ discretization (Model 1)
\cite{Campbell}, two energy-conserving discretizations, namely
the Speight-one (Model 2) \cite{SpeightKleinGordon} and the one
of Ref. \cite{Saxena} also labeled CKMS hereafter (Model 3), and two
momentum-conserving discretizations stemming from Ref.
\cite{KevrekidisPhysD}, labeled K1 (Model 4) and K2 (Model 5),
respectively. For each one of these models, we begin by examining
the properties of the fundamental building block nonlinear wave
solutions, namely the discrete kinks (and antikinks). We show how
to obtain such solutions analytically or semi-analytically and
subsequently examine the spectrum of small-amplitude
excitations (linearization) around them, among other properties
(such as stability) because this spectrum plays a nontrivial role
in the outcome of wave interactions. Finally, we focus on the
latter (i.e., on solitary wave collisions between kinks and
antikinks) and attempt to extract salient features of such
interactions as a function of the lattice spacing $h$ and for a
set of different speeds (and also different initial separations
between the kinks).

Our main findings can be summarized as follows:

\begin{itemize}
\item The different models yield kink profiles that are different between
them and from the continuum limit. These differences are strongest
for the CKMS model and are found to lead to shrinking kinks for
the energy conserving models 1-3, while they lead to expanding
kinks in models 4-5. The deviation from the relevant continuum
profile grows as $h^2$.
\item The boosting of the kinks in order to induce their collision
excites their internal modes. This plays a significant role in the
collisions, since for different initial distances, the excitation
of the internal mode will carry a different phase (at the moment
of collision) and may accordingly lead to different collision
outcomes.
\item The different models have different properties as regards
the elasticity of their collisions. The most inelastic collisions
occur in the standard discretization of model 1. Perhaps the next
least elastic collisions occur in the CKMS Model 3, then K1 (Model
4), Speight (Model 2) and K2 (Model 5) in order of increasing
elasticity.
\item The elasticity of collisions changes as a function of the
lattice spacing. In fact, remarkably so, the collisions are {\it
more elastic for larger values of the lattice spacing}. This is
also demonstrated in the decreasing dependence of the critical
velocity (beyond which the solitary waves separate after one
collision) as a function of $h$.
\end{itemize}

The presentation of our results will be structured as follows. In
section \ref{sec:DisreteModel}, we will present the various models
and, in section \ref{sec:KinkDiscussion}, compare their kink
solutions and spectral properties. In section \ref{sec:Numerics},
we will examine the properties of the collisions of the different
models focusing on a few typical speeds of the incoming waves for
different initial distances and for different initial lattice
spacings. Finally, in section \ref{sec:Conclusions}, we will
summarize our findings and present our conclusions, as well as
some open questions for future study.

%--------------------------------------------------------------------------
\section{ $\phi^4$ field and its various discretizations } \label{sec:DisreteModel}
%--------------------------------------------------------------------------

Starting from the continuum limit of the model, we note that the
one-dimensional $\phi^4$ field is described by the Lagrangian
${\cal L}=E_K-E_P$ with the kinetic and potential energy
functionals defined, respectively, by
\begin{equation} \label{KinEn}
   E_K =  \frac{1}{2} \int_{-\infty}^{\infty} \phi_{t}^2dx\,,
\end{equation}
\begin{equation} \label{PotEn}
   E_P = \frac{1}{2}\int_{-\infty}^{\infty} \left[ \phi_{x}^2
   +\left(1 - \phi^2 \right)^2 \right]dx\,,
\end{equation}
where $\phi(x,t)$ is the scalar field of interest and subscript
indices mean partial derivatives with respect to the corresponding
variable. The resulting Euler-Lagrange equation is obtained by
demanding that $\phi$ be a local extremum of the action $S=\int
{\cal L} {\rm d}t$, and it reads
\begin{equation} \label{phi4}
   \phi_{tt} = \phi_{xx} +2\phi(1 - \phi^2)\,.
\end{equation}

The following kink (antikink) solution to Eq. (\ref{phi4}),
\begin{eqnarray} \label{phi4kink}
\phi(x,t)= \pm \tanh\frac{x - x_{0} - vt}{\sqrt{1-v^2}},
\end{eqnarray}
is one of the simplest examples of topological solitons. In Eq.
(\ref{phi4kink}), $v$ is the kink velocity and $x_0$ is its
arbitrary initial position (signalling the translational
invariance of the continuum model discussed in the previous
section).

The first integral of the static version of Eq. (\ref{phi4}),
\begin{equation} \label{FI}
   U(x) \equiv \phi_{x}^2 -(1 - \phi^2)^2=0\,,
\end{equation}
plays an important role in our considerations. The integration
constant was set to zero in Eq. (\ref{FI}), which is sufficient
for obtaining the kink solutions. The first integral can also be
taken in a modified form, e.g., as
\begin{equation} \label{FImodif}
   u(x) \equiv \pm \phi_{x} - 1 + \phi^2=0\,.
\end{equation}

We study various lattice dynamical equations obtained by
discretizing Eq. (\ref{phi4}) on the lattice $x=nh$, where
$n=0,\pm 1, \pm 2 ...$, and $h$ is the lattice spacing. The
general form of the lattice equations studied herein is
\begin{eqnarray} \label{GeneralDiscrete}
   \ddot{\phi}_{n}&=& \Delta_2\phi_n
   + F(\phi_{n-1},\phi_{n},\phi_{n+1})\nonumber \\
   &\equiv& D(\phi_{n-1},\phi_{n},\phi_{n+1}),
\end{eqnarray}
where
\begin{eqnarray} \label{SecDerivDiscr}
   \Delta_2\phi_n=\frac{1}{h^2}(\phi_{n-1} -2\phi_{n} +\phi_{n+1}),
\end{eqnarray}
and, in the continuum limit $(h \rightarrow 0)$,
\begin{equation} \label{ContLimit}
   F(\phi_{n-1},\phi_{n},\phi_{n+1}) \rightarrow 2\phi(1-\phi^2).
\end{equation}

As mentioned previously, of particular interest will be the
lattices whose static solutions, satisfying the three-point static
problem corresponding to Eq. (\ref{GeneralDiscrete}),
\begin{equation} \label{ThreePointStatic}
   D(\phi_{n-1},\phi_{n},\phi_{n+1}) = 0,
\end{equation}
can be found from a reduced two-point problem of the form
\begin{equation} \label{TwoPointStatic}
   U(\phi_{n-1},\phi_{n}) = 0,
\end{equation}
or of the form
\begin{equation} \label{TwoPointStaticModif}
   u(\phi_{n-1},\phi_{n}) = 0.
\end{equation}
Equations (\ref{TwoPointStatic}) and (\ref{TwoPointStaticModif})
are the discretized first integrals (DFIs) obtained by
discretizing Eq. (\ref{FI}) and Eq. (\ref{FImodif}), respectively
\cite{JPhysA} (see also \cite{Barashenkov}).

Lattices whose static solutions can be found from the two-point
DFI are called translationally invariant because the equilibrium
solution can be obtained iteratively from the nonlinear algebraic
equation starting from an arbitrary admissible value $\phi_n$. In
other words, translationally invariant lattices support a continuum
rather than a discrete set of equilibrium solutions parametrized,
e.g., by the position of the $n$th particle, $\phi_n$.

Let us now calculate the PN barrier in the models whose static
solutions can be found from a two-point DFI, i.e., in the
translationally invariant lattices. We consider a continuum set
of equilibrium solutions parameterized by the position of the $n$th
particle, $\phi_n$, in the range $\phi_n \in
[\phi_n^{(1)},\phi_n^{(2)}]$, assuming that all values of $\phi_n$
within this range are admissible.

The work done by the inter-particle and the background forces
(originating from the discretized background potential) to move
(quasi-statically) the $n$th particle from the configuration
$\phi^{(1)}$ to the configuration $\phi^{(2)}$ is
\begin{equation} \label{Wn}
   W_n= \int_{\phi_n^{(1)}} ^{\phi_n^{(2)}}
   D(\phi_{n-1},\phi_{n},\phi_{n+1}) {\rm d}\phi_n,
\end{equation}
and the total work performed to ``transform'' the whole chain from
$\phi^{(1)}$ to $\phi^{(2)}$ is
\begin{equation} \label{W}
   W=\sum_{n=-\infty}^{\infty}W_n.
\end{equation}
However, in the translationally invariant models, the availability
of a path of equilibrium configurations allowing to transit from
$\phi^{(1)}$ to $\phi^{(2)}$ leads to $D=0$ and thus, $W_n=0$ for
all $n$. This, in turn, results in $W=0$. This result suggests
that there is no energy cost to transform quasi-statically one
equilibrium solution into another through a continuous set of
equilibrium solutions. In other words, the height of the
Peierls-Nabarro barrier {\em calculated along this path} is zero.
For Hamiltonian lattices, the total work is path-independent
(and equal to the potential energy difference between the final
and initial state)
and we can claim that, in such dynamical lattices,
the PN potential is zero. For non-Hamiltonian lattices the
work is path-dependent and we can only claim the absence of the PN
barrier along the path considered above (which, however, is a
natural one). While there are mathematical subtleties as regards
whether this notion yields zero PNb more generally
for translationally invariant lattices, this is the definition of
PNb-free models that will be used herein. A more detailed
examination of the, admittedly interesting, pertinent topics is
outside the scope of the present study focusing on the comparison
between different discrete $\phi^4$ models and will be delegated
to a future publication.

In the following, we consider various discrete $\phi^4$ models
reported in the literature describing their kink solutions, their
spectra of small-amplitude vibrations around vacuum solutions
($\phi_n=\pm1$), and also the spectra of lattices containing one
static kink, revealing the kink's internal vibrational modes.
Physical quantities conserved by the lattices are given, if they
exist. These results will be quite relevant also in the discussion
of kink-antikink collision outcomes.

%--------------------------------------------------------------------------
\subsection{Classical discretization (model 1)} \label{sec:Model1}
%--------------------------------------------------------------------------

The ``standard''  discretization of Eq. (\ref{phi4}) is
\cite{Campbell}
\begin{equation} \label{Classical}
   \ddot{\phi}_{n}= \Delta_2\phi_n
   + 2\phi_{n}(1 - \phi_{n}^2),
\end{equation}
and this is the only lattice in this study that possesses a
Peierls-Nabarro barrier.

Model 1 conserves the Hamiltonian (total energy)
\begin{eqnarray} \label{H1}
   H_1= \frac{h}{2} \sum_n \left[ \dot{\phi}_n^2
   + \frac{\left(\phi_{n+1}-\phi_n\right)^2}{h^2}
   +\left(1 - \phi_n^2 \right)^2  \right].
\end{eqnarray}
Static kink solutions in model 1 exist only for those waves centered at
a lattice site (unstable) or in the middle between two neighboring
lattice sites (stable). Solutions can be found by various
numerical techniques. As a first approximation, one can adopt Eq.
(\ref{phi4kink}) to write the following {\em approximate} static
kink solution

\begin{eqnarray} \label{phi4kinkApprox}
\phi_n= \pm \tanh [h(n - x_{0})].
\end{eqnarray}

One can use this {\it ansatz} in a fixed point scheme (such as a
Newton method) for $x_0=0$ or $x_0=1/2$ (mod 1), to identify the
exact discrete static solutions $\phi_n^0$.

\begin{figure}
%\begin{center}
\includegraphics{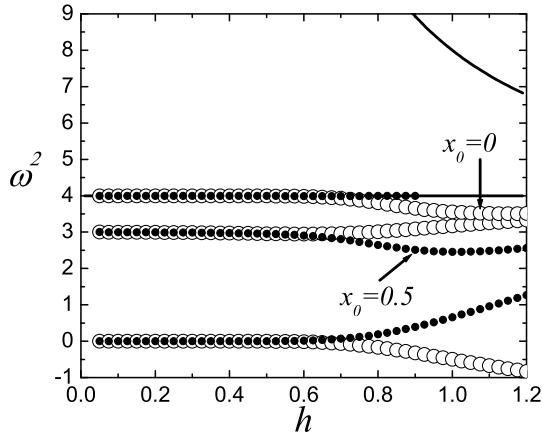}
%\end{center}
\caption{Model 1, Eq. (\ref{Classical}): Frequencies of the kink's
internal modes for different magnitudes of the discreteness
parameter $h$. Results for the on-site (circles) and inter-site
(dots) kinks. Two solid lines show the borders of the spectrum of
vacuum, Eq. (\ref{SpecVacClassic1}). The on-site kink is unstable because
the spectrum contains imaginary frequencies, while the inter-site kink
is stable. All three internal models are below the phonon band.}
\label{fig1}
\end{figure}

Subsequently, introducing the ansatz
$\phi_n(t)=\phi_n^0+\varepsilon_n(t)$ (where $\phi_n^0$ is an
equilibrium solution and $\varepsilon_n(t)$ is a small
perturbation), we linearize Eq. (\ref{Classical}) with respect to
$\varepsilon_n$ and obtain the following equation:
\begin{eqnarray} \label{LinClassic1}
   \ddot{\varepsilon}_n=\Delta_2\varepsilon_{n} + 2\varepsilon_{n}
   -6 (\phi_n^0)^2 \varepsilon_{n}.
\end{eqnarray}
For the small-amplitude phonons, $\varepsilon_{n}=\exp(i k n +i
\omega t)$, with frequency $\omega$ and wave number $k$, Eq.
(\ref{LinClassic1}) is reduced to the following dispersion
relation:
\begin{eqnarray} \label{SpecClassic1}
   \omega^2=\frac{4}{h^2}\sin^2\left( \frac{k}{2} \right)
   -2 + 6(\phi_n^0)^2.
\end{eqnarray}
From Eq. (\ref{SpecClassic1}), the spectrum of the vacuum
solution, $\phi_n^0=\pm 1$, is
\begin{eqnarray} \label{SpecVacClassic1}
   \omega^2=4+\frac{4}{h^2}\sin^2\left( \frac{k}{2} \right).
\end{eqnarray}
The spectrum of the lattice when linearizing around
 a static kink is shown in Fig.
\ref{fig1}.

%--------------------------------------------------------------------------

\subsection{ Energy-conserving model 2 }\label{sec:Model2}

%--------------------------------------------------------------------------

Here we use the following DFI obtained from Eq.
(\ref{TwoPointStaticModif}),
\begin{eqnarray} \label{SpeightKink}
   u_2 \equiv \pm\frac{\phi_{n}-\phi_{n-1}}{h}  %\nonumber \\
   - 1+ \frac{\phi_{n-1}^2
   +\phi_{n-1}\phi_{n}+\phi_{n}^2}{3} = 0.
\end{eqnarray}
The Hamiltonian, ${\cal H}=E_K+E_P$, defined by Eq. (\ref{KinEn})
and Eq. (\ref{PotEn}), can be discretized as follows,
\begin{eqnarray} \label{H2}
   H_2= h \sum_n \left( \frac{\dot{\phi}_n^2}{2}
   + u_2^2 \right),
\end{eqnarray}
which gives the equations of motion of the energy-conserving model
after Speight \cite{SpeightKleinGordon} (see also
\cite{SpeightPhi4}),
\begin{eqnarray} \label{Speight}
   \ddot{\phi}_n&=&
   -u_2(\phi_{n-1},\phi_n)\frac{\partial }{{\partial \phi_n}}
  u_2(\phi_{n-1},\phi_n) %\nonumber \\
  -u_2(\phi_{n},\phi_{n+1})\frac{\partial }{{\partial \phi_n}}
  u_2(\phi_{n},\phi_{n+1}) \nonumber \\
   &=&\left(1 + \frac{h^2}{3} \right) \Delta_2\phi_n+2\phi_n %\nonumber \\
   -\frac{1}{9}\left[2\phi_n^3+(\phi_n+\phi_{n-1})^3
   +(\phi_n+\phi_{n+1})^3\right].
\end{eqnarray}
It is clear that the static solutions to Eq. (\ref{Speight}) can
be found from the two-point problem, Eq. (\ref{SpeightKink}). We
have
\begin{eqnarray} \label{SpeightKinksolution1}
   \phi_{n\pm 1}=-\frac{\phi_{n}}{2}\mp \frac{3}{2h}
   \pm \frac{\sqrt{3}}{2} \sqrt{-\phi_{n}^2
   \pm \frac{6}{h}\phi_{n}+ \frac{3}{h^2}+4},
\end{eqnarray}
where one can take either the upper or the lower signs. The kink
solution can be obtained iteratively from Eq.
(\ref{SpeightKinksolution1}), starting from any $|\phi_n|<1$. For
the on-site and inter-site kinks one should take for the initial
value $\phi_n=0$ and $\phi_n=3/h-\sqrt{3+9/h^2}$, respectively.

The equation of motion, Eq. (\ref{Speight}), linearized in the
vicinity of an equilibrium solution $\phi_n^0$ yields
\begin{eqnarray} \label{LinSpeight1}
   \ddot{\varepsilon}_n=\left( 1+\frac{h^2}{3} \right)
   \Delta_2\varepsilon_{n} + 2\varepsilon_{n}  %\nonumber \\
   -\frac{1}{9} \Big[ 6(\phi_n^0)^2 \varepsilon_{n}
   +3(\phi_n^0+\phi_{n-1}^0)^2 (\varepsilon_{n}+\varepsilon_{n-1}) %\nonumber \\
   +3(\phi_n^0+\phi_{n+1}^0)^2 (\varepsilon_{n}+\varepsilon_{n+1})
   \Big].
\end{eqnarray}

The spectrum of the vacuum solution, $\phi_n^0=\pm 1$, is
\begin{equation}\label{Spec2}
   \omega^2=4+4\frac{1-h^2}{h^2}\sin^2\left(\frac{k}{2}\right) .
\end{equation}

\begin{figure}
\includegraphics{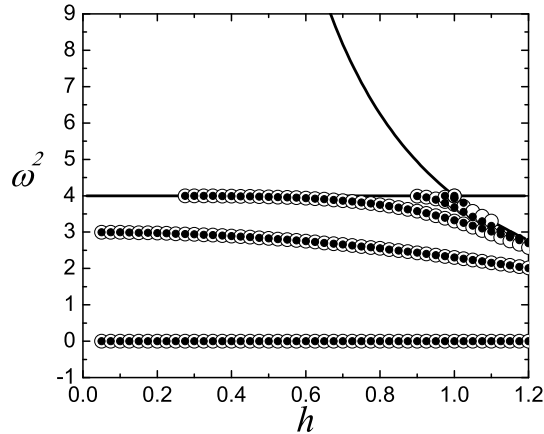}
\caption{Model 2, Eq. (\ref{Speight}). Same as in Fig. \ref{fig1}.
Notice the difference of the mode closest to the origin which in
this case remains at $\omega^2=0$ contrary to what is the case for
model 1. There are three internal modes. However, note that a fourth
mode appears for $h >0.85$. There is a zero mode for all values of $h$
indicating the absence of a  PN barrier.} \label{fig2}
\end{figure}

On the other hand, the spectrum of linearization around a kink is
shown in Fig. \ref{fig2}.

%--------------------------------------------------------------------------

\subsection{ Energy-conserving model 3 }\label{sec:Model3}

%--------------------------------------------------------------------------

We take the DFI, corresponding to Eq. (\ref{FI}), in the form
\begin{eqnarray} \label{DFI3}
U_3 \equiv \frac{1}{h^2}(\phi_n-\phi_{n-1})^2-
(1-\phi_{n-1}\phi_{n})^2=0.
\end{eqnarray}

The equations of motion of the model of CKMS \cite{Saxena},
\begin{eqnarray} \label{Saxena}
   \ddot{\phi}_n  &=& \frac{U_3(\phi_n,\phi_{n+1}) - U_3(\phi_{n-1},\phi_n) }
   {(\phi_{n+1}-\phi_{n-1}) (1-h^2\phi_n^2)} \nonumber \\
   &=&\Delta_2\phi_n + 2\frac{\phi_n -
   \phi_n^3}{1 -  h^2 \phi_n^2},
\end{eqnarray}
can be obtained from the Hamiltonian
\begin{eqnarray}
H_3 =\frac{1}{2}\sum_{n} \Big[ {\dot \phi}_n^2 +
\frac{(\phi_n-\phi_{n-1})^2}{h^2}+ V(\phi_n)\Big], \label{calh}
\end{eqnarray}
where the potential $V(\phi_n)$ is given by
\begin{eqnarray}
V(\phi_n)=-\frac{1}{h^2} \left(\phi_n^2 + \frac{1-h^2}{h^2}\ln
\left|\phi_n^2 - \frac {1}{h^2}\right| \right). \label{HamSaxena}
\end{eqnarray}
The exact static kink (antikink) solution is \cite{Saxena}
\begin{eqnarray} \label{KinkCooper}
   \phi_n=\pm\tanh [\beta h(n - x_0)], \quad \tanh(\beta h)=h,
\end{eqnarray}
where $x_0$ is the arbitrary position of the solution.

Alternatively, the kink solution can be found from Eq.
(\ref{DFI3}). We come to the iterative formula,
\begin{eqnarray} \label{KinkCooper2}
   \phi_n=\frac{\phi_{n-1} \pm h}{1\pm
   h \phi_{n-1}},
\end{eqnarray}
where one can choose either the upper or the lower signs and one
can interchange $\phi_n$ and $\phi_{n-1}$. To obtain a kink
centered on a lattice site, one should use as a starting point the
value
$\phi_n=0$, while for a kink centered in the middle between two
neighboring sites, $\phi_n=1/h-\sqrt{1/h^2-1}$.

The linearized equation of motion reads
\begin{eqnarray}
\ddot{\varepsilon}_n=\Delta_2\varepsilon_{n} +2\frac {1+\left(h^2
-3 \right)(\phi_n^0)^2 + h^2 (\phi_n^0)^4} {\left[ 1- h^2
(\phi_n^0)^2 \right]^2} \varepsilon_{n}. \label{LinSaxena}
\end{eqnarray}
The spectrum of vacuum solutions $\phi_n^0=\pm 1$ is
\begin{eqnarray}
\omega^2=\frac{4}{1-h^2 } + \frac{4}{h^2} \sin^2 \left(\frac{k}{2}
\right). \label{SpecVacSaxena}
\end{eqnarray}

\begin{figure}
\includegraphics{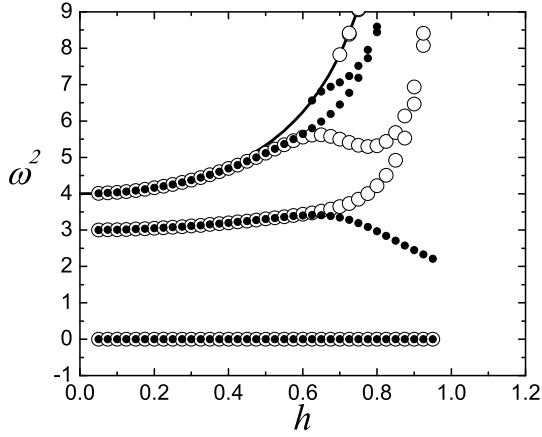}
\caption{Model 3, Eq. (\ref{Saxena}): Same as in Fig. \ref{fig1}.
The vacuum solution $\phi_n=\pm 1$ is unstable for $h>1$. The upper
edge of the phonon spectrum lies above the scale of the figure. There
are three internal modes. A fourth mode appears for $h > 0.6$. Note
that there is a zero mode for all values of $h$ indicating the absence
of a PN barrier.}
\label{fig3}
\end{figure}

On the other hand, the spectrum of
the CKMS lattice with a kink is shown in Fig. \ref{fig3}.

%--------------------------------------------------------------------------
\subsection{Momentum-conserving model 4 }\label{sec:Model4}
%--------------------------------------------------------------------------

Discretizing Eq. (\ref{FI}) as follows,
\begin{eqnarray} \label{DFI4}
U_4 \equiv \frac{1+h^2}{h^2}(\phi_n-\phi_{n-1})^2-
(1-\phi_{n-1}\phi_{n})^2=0,
\end{eqnarray}
we come to the model reported in the work of
\cite{KevrekidisPhysD} (motivated by its corresponding, so-called
Ablowitz-Ladik sibling discretization for the nonlinear
Schr{\"o}dinger equation \cite{AL})
\begin{eqnarray} \label{PhysD4}
   {\ddot \phi_n}&=& \frac{U_4(\phi_{n},\phi_{n+1})
   -U_4(\phi_{n-1},\phi_{n})}{\phi_{n+1}-\phi_{n-1}} \nonumber \\
   &=&\Delta_2\phi_n
   +(\phi_{n+1}+\phi_{n-1})(1-\phi_{n}^2) .
\end{eqnarray}
This non-Hamiltonian PNb-free model conserves the momentum
\cite{KevrekidisPhysD} which has the form:
\begin{eqnarray} \label{mom4}
P_4=\sum_n \dot{\phi}_n (\phi_{n+1}-\phi_{n-1}).
\end{eqnarray}

The exact static kink (antikink) solution is
\begin{eqnarray} \label{KinkCooper4}
   \phi_n=\pm\tanh [\beta h(n - x_0)], \quad
   \tanh(\beta h)=\frac{h}{\sqrt{1+h^2}},
\end{eqnarray}
where $x_0$ is the arbitrary position of the solution.

Alternatively, the kink solution can be found iteratively from
\begin{eqnarray} \label{KinkCooper42}
   \phi_n=\frac{\phi_{n-1} \pm h/\sqrt{1+h^2}}{1\pm
   \phi_{n-1}h/\sqrt{1+h^2}},
\end{eqnarray}
where one can choose either the upper or the lower signs and one
can interchange $\phi_n$ and $\phi_{n-1}$. To obtain the on-site
(inter-site) kink one should use as initial value $\phi_n=0$
$(\phi_n=\sqrt{1+h^2}/h-1/h)$.

The equation of motion, Eq. (\ref{PhysD4}), linearized in the
vicinity of an equilibrium solution $\phi_n^0$ assumes the form
\begin{eqnarray} \label{LinPhysD4}
   \ddot{\varepsilon}_n=\left( 1+h^2 \right)
   \Delta_2\varepsilon_{n} + 2\varepsilon_{n} % \nonumber \\
   -(\phi_n^0)^2( \varepsilon_{n-1}+ \varepsilon_{n+1})
   -2\phi_n^0(\phi_{n-1}^0+\phi_{n+1}^0) \varepsilon_{n}.
\end{eqnarray}

The spectrum of vacuum, $\phi_n^0=\pm 1$, coincides with that of
model 1, Eq. (\ref{SpecVacClassic1}).

\begin{figure}
\includegraphics{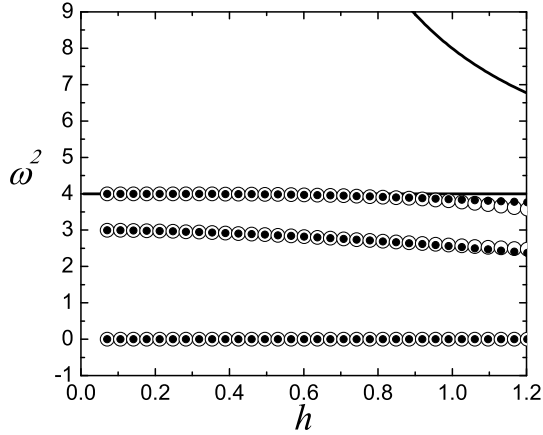}
\caption{Model 4, Eq. (\ref{PhysD4}). Same as in
  Fig. \ref{fig1}. There are three internal modes including the
  $\omega^2 = 0$ mode indicating the absence of a PN barrier.}
\label{fig4}
\end{figure}

However, the spectrum of the linearization around a kink is different
as shown in
Fig. \ref{fig4}.

%--------------------------------------------------------------------------
\subsection{ Momentum-conserving model 5 }\label{sec:Model5}
%--------------------------------------------------------------------------

Discretizing Eq. (\ref{FI}) as
\begin{eqnarray} \label{DFI5}
   U_5 \equiv \frac{1}{h^2}(\phi_n-\phi_{n-1})^2
   -1 +\phi_{n-1}^2 +\phi_{n}^2 %\nonumber \\
   \frac{1}{4}(\phi_{n-1}^4 +\phi_{n}^4)
   -\frac{1}{2}\phi_{n-1}^2\phi_{n}^2=0,
\end{eqnarray}
we obtain another momentum-conserving model of the type of
\cite{KevrekidisPhysD} (see also \cite{Barashenkov})
\begin{eqnarray} \label{PhysD5}
   {\ddot \phi_n}&=& \frac{U_5(\phi_{n},\phi_{n+1})
   -U_5(\phi_{n-1},\phi_{n})}{\phi_{n+1}-\phi_{n-1}} \nonumber \\
   &=&\Delta_2\phi_n+\frac{\phi_{n+1}+\phi_{n-1}}{4}
   (4-\phi_{n-1}^2-2\phi_{n}^2-\phi_{n+1}^2) .
\end{eqnarray}

This non-Hamiltonian PNb-free model conserves the momentum of Eq.
(\ref{mom4}).

Static solutions in this model can be found iteratively by solving
the quartic Eq. (\ref{DFI5}).

The equation of motion, Eq. (\ref{PhysD5}), linearized in the
vicinity of an equilibrium solution $\phi_n^0$ is
\begin{eqnarray} \label{LinPhysD5}
   \ddot{\varepsilon}_n&=&
   \Delta_2\varepsilon_{n} + \varepsilon_{n-1}
   + \varepsilon_{n+1}  \nonumber \\
   &-&\frac{\phi_{n-1}^0+\phi_{n+1}^0}{2}
   ( \phi_{n-1}^0\varepsilon_{n-1} + 2\phi_{n}^0\varepsilon_{n}
   + \phi_{n+1}^0\varepsilon_{n+1})  \nonumber \\
   &-&\frac{\varepsilon_{n-1} + \varepsilon_{n+1}}{4}
   \left[(\phi_{n-1}^0)^2+2(\phi_{n}^0)^2+(\phi_{n+1}^0)^2\right].
\end{eqnarray}

The spectrum of vacuum is the same as for model 2, Eq.
(\ref{Spec2}).

\begin{figure}
\includegraphics{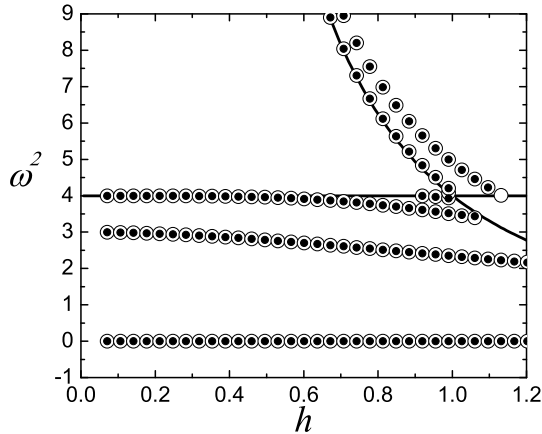}
\caption{Model 5, Eq. (\ref{PhysD5}). Same as in Fig. \ref{fig1}.
A unique feature of this discretization is the presence of the
vibrational modes lying above the phonon spectrum. There are three
internal modes (including the zero mode) and three additional
modes for higher values of $h$ indicating the presence of modes above
the phonon spectrum.} \label{fig5}
\end{figure}

Furthermore, the spectrum of the linearization around a kink is
shown in Fig. \ref{fig5}.

%--------------------------------------------------------------------------
\section{Comparison of Kink Properties}\label{sec:KinkDiscussion}
%--------------------------------------------------------------------------

%--------------------------------------------------------------------------
\subsection{Spectra of vacuum and kink's internal modes}\label{sec:Discussion}
%--------------------------------------------------------------------------

We have presented the spectra for the classical $\phi^4$ model (model
1) and for the four models free of the Peierls-Nabarro barrier
(models 2-5). All models share the same continuum limit; that is
why, for small $h$, their properties are close and they only start to
deviate from each other, as $h$ increases.

If we divide the models in groups by the quantities they
conserve, then models 1-3 belong to the energy-conserving group while
models 4 and 5 conserve the momentum of Eq. (\ref{mom4}).

Models 3 and 4 have the static solutions derived in \cite{Saxena}
and \cite{DKYF}. Comparing the DFIs of these models, Eq.
(\ref{DFI3}) and Eq. (\ref{DFI4}), we can see that the solutions
for model 4 can be obtained from those for model 3 by substituting
$h \rightarrow h/\sqrt{1+h^2}$. Exact static kink solutions are
given for model 3 by Eq. (\ref{KinkCooper}) or Eq.
(\ref{KinkCooper2}) and for model 4 by Eq. (\ref{KinkCooper4}) or
Eq. (\ref{KinkCooper42}).

Exact static kink solutions for model 2 can be found iteratively
from Eq. (\ref{SpeightKinksolution1}), while  the ones of model 5
can be obtained by
solving the quartic Eq. (\ref{DFI5}). For Model 1, the full
3-point problem of Eq. (\ref{ThreePointStatic}) needs to be solved.

Comparing the spectra of the vacuum (band edges of the spectra are
shown by solid lines in Figs. \ref{fig1} - \ref{fig5}), we note
that:

\begin{itemize}
\item Model 4 has the same spectrum of vacuum as the classical model 1,
and the width of this spectrum vanishes only when $h\rightarrow
\infty$. The vacuum solution is always stable because $\omega^2>0$
for any $h$.
\item Models 2 and 5 have the same spectrum of the vacuum. The width of
the spectrum vanishes at $h=1$. Close to this value of $h$, the
phonon spectrum is narrow and hence potential phonon radiation (of
a kink-like structure due to resonance of internal mode harmonics
with the phonon band) is minimized. The vacuum solution is always
stable because $\omega^2>0$ for any $h$.
\item Model 3 has an $h$-dependent cubic term; that is why the
lower boundary of the spectrum is also $h$-dependent, while in all other
models it is constant ($\omega^2=4$). In this model, the vacuum
solution is stable only for $0<h<1$.

\end{itemize}

Subsequently, examining the spectra of lattices containing a
static kink, we note that (frequencies of kink's internal modes
are shown in Figs. \ref{fig1} - \ref{fig5} by circles and dots):

\begin{enumerate}
\item Models 2-5 are PNb-free because they have a zero-frequency
mode, which is the, so-called, translational (or the Goldstone)
mode of the kink. For model 1, the corresponding mode has a
non-zero frequency (in fact, it depends on $h$ as $\propto
\exp(-\pi^2/h)$; see e.g., \cite{pla}), signalling the presence of
the Peierls-Nabarro barrier. Given the form of its $h$-dependence,
for small $h$ $(<0.4)$, even for model 1 this mode has a nearly
zero frequency (see Fig. \ref{fig1}). This is the weakly perturbed
translational mode of the continuum $\phi^4$ equation.
\item For small $h$ $(<0.4)$ and even for moderate $h$ $(<0.8)$, for all
five models, apart from the translational mode we have two kink
internal modes lying below the phonon spectrum, one of them very
close to the edge of the phonon spectrum ($\omega^2=4$) and
another one in the vicinity of $\omega^2 \sim 3$ (i.e., the
corresponding continuum limit of this mode \cite{sugiyama}).
%For
%small $h$, the mode close to the bottom edge is in fact extremely
%wide, it vanished on the distances much greater than the kink
%width.
Additional internal modes may emerge for large $h$, but we
will focus on smaller values of $h$ (i.e., for $h<0.5$) in the
collision results that follow, hence we do not discuss these
further here.
\item Model 5, in contrast to all other models, is the model with
internal modes lying not only below but also above the phonon
band. Such modes, in contrast to the previously reported ones below the band,
are of the short-wave (staggered) type and, for this reason,
their excitation (or lack thereof) can be sensitive to the
position of the collision point with respect to the lattice.
\end{enumerate}

%--------------------------------------------------------------------------
\subsection{ Static kink profile and kink boosting}\label{sec:StaticKink}
%--------------------------------------------------------------------------

\begin{figure}
\includegraphics{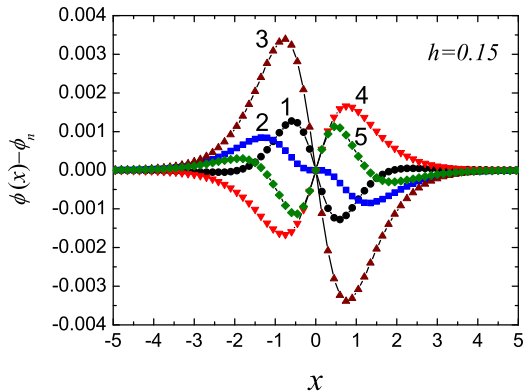}
\caption{Difference between the continuum and the discrete static
kink profiles in the five models at $h=0.15$. The continuum static
kink is given by Eq. (\ref{phi4kink}) with $v=0$. The kink in
model 3 has the largest deviation from the continuum kink profile.
Kinks in the energy-conserving discrete models (1 to 3) have a width
smaller than the continuum kink while for the momentum-conserving
models 4 and 5, the situation is reversed.}
\label{KinkInitialShape}
\end{figure}

\begin{figure}
\includegraphics{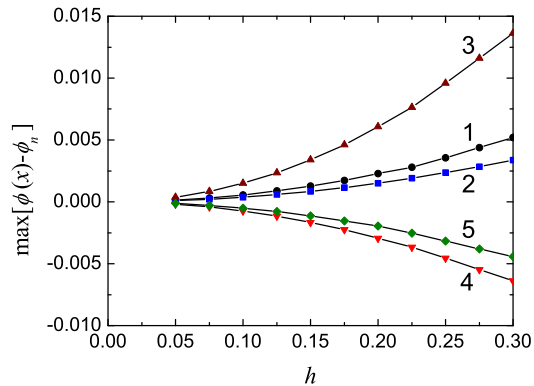}
\caption{The maximal difference between the continuum and the
discrete static kink profiles shown as a function of the spacing
$h$ for the five models. The amplitude of the deviation from the
continuum kink profile increases with the discreteness parameter
as $h^2$.} \label{MaxKinkShape}
\end{figure}

It is also of interest to compare the static kink profiles in the
five discrete models (to examine the relevant deviations between
them and the continuum limit from which they are derived). In Fig.
\ref{KinkInitialShape} we present the difference between the static kink
profiles of the different models and that of the continuum static kink, Eq.
(\ref{phi4kink}). The lattice spacing is $h=0.15$. It is clear
that the kink in the CKMS model 3 has the largest deviation from
the continuum kink profile. Kinks in the energy-conserving
discrete models (1 to 3) have widths which are smaller than that
of the continuum kink while for the momentum-conserving models 4
and 5 the situation is reversed. Furthermore, in Fig.
\ref{MaxKinkShape}, we show how this difference between continuum
and discrete kink is amplified as $h$ is increased. This is carried out
by showing the maximal difference between the two kinks as an
$h$-dependent diagnostic which reveals that the relevant
difference grows as $h^2$ as $h$ increases.

\begin{figure}
\includegraphics{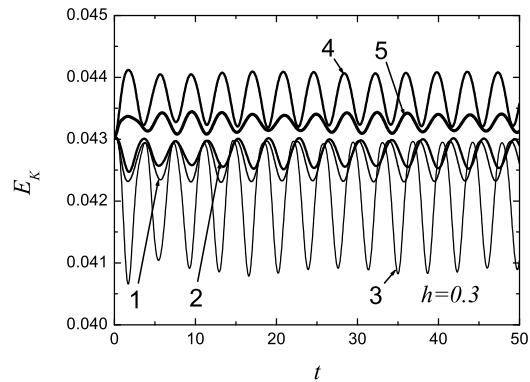}
\caption{Kinetic energy of a lattice containing a kink boosted at
$t=0$ with $v=0.25$ with the help of Eq. (\ref{KinBoost}) for the
five discrete models. The lattice spacing is $h=0.3$.} \label{fig6}
\end{figure}

In order to induce collisions, the kink needs to be set into
motion in the discrete system. This can be achieved in a variety
of ways. Here we have used the most standard one, namely Lorentz
boosting the static kink to speed $v$, according to the continuum
{\it ansatz}:
\begin{eqnarray} \label{KinBoost}
\phi(x,t)= \pm \tanh\frac{h(n - x_{0}) - vt}{\sqrt{1-v^2}}.
\end{eqnarray}

The results presented previously about the static kink also have a
direct bearing on the kink boosting. In Fig. \ref{fig6} we show
the kinetic energy of a lattice containing a kink boosted at $t=0$
with velocity $v=0.25$ through Eq. (\ref{KinBoost}). Lines of
different thickness show the results for the five discrete models.
The lattice spacing is $h=0.3$ in this figure (even though
similar, yet less pronounced results have been obtained for
smaller $h$; again the relevant trend is quadratic in $h$).
$E_{K}(t)$ oscillates with the frequency close to the kink
internal mode frequency of $\omega^2\sim 3$. At $t=0$ we have
$E_{K}=0.043$ and for the energy-conserving models 1 to 3,
$E_{K}(t)$ is below this value, while for the momentum-conserving
models 4 and 5, it is above this value. This is in sync with the
static results where it was shown that models 4 and 5 have a
correction to the continuum kink profile of opposite sign than the
models 1 to 3 (see also Fig. \ref{KinkInitialShape}).  Model 3 shows
the largest amplitude of kinetic energy oscillations, again in
agreement with the static results.

We have investigated another boosting method that uses the
dynamical solution of the form
$\phi_n(t)=\phi_n^0+vt\varepsilon_n$, where $\phi_n^0$ is the
static kink solution, $\varepsilon_n$ is the normalized
translational kink's internal mode corresponding to the multiple
eigenvalue $\omega^2=0$, and $v$ is the amplitude that plays the
role of kink's velocity. We found this method to be very good
(internal modes were not excited for any $h$) for small $v$, as it
should be, because the accuracy of the linearized equations of
motion increases as the eigenmode amplitude decreases.
However, for velocities of order of $v\sim 0.1$ the accuracy
of this method is insufficient because it does not take into
account the Lorentz correction of kink's width. The ansatz Eq.
(\ref{KinBoost}) takes into account this correction, but it does
not take into account the discreteness of media and, hence, naturally
it is less accurate for large $h$.

The best results were obtained for the use of Eq. (\ref{KinBoost})
together with the addition of the kink's internal mode with the
amplitude chosen to compensate the excitation of such mode. In the
present study we did not use this more complicated/fine tuned method.

As a result of these considerations, even for relatively small
$h$, the kink boosted employing Eq. (\ref{KinBoost}) carries an
internal mode of non-vanishing amplitude. This internal mode often
plays  a nontrivial role in determining the outcome of the
collision in what follows.

%--------------------------------------------------------------------------
\section{Collision Results}\label{sec:Numerics}
%--------------------------------------------------------------------------

%--------------------------------------------------------------------------
\subsection{Numerical findings for different lattice spacings, initial
speeds and kink-antikink separations}
%--------------------------------------------------------------------------

\begin{figure}
%\begin{center}
\includegraphics{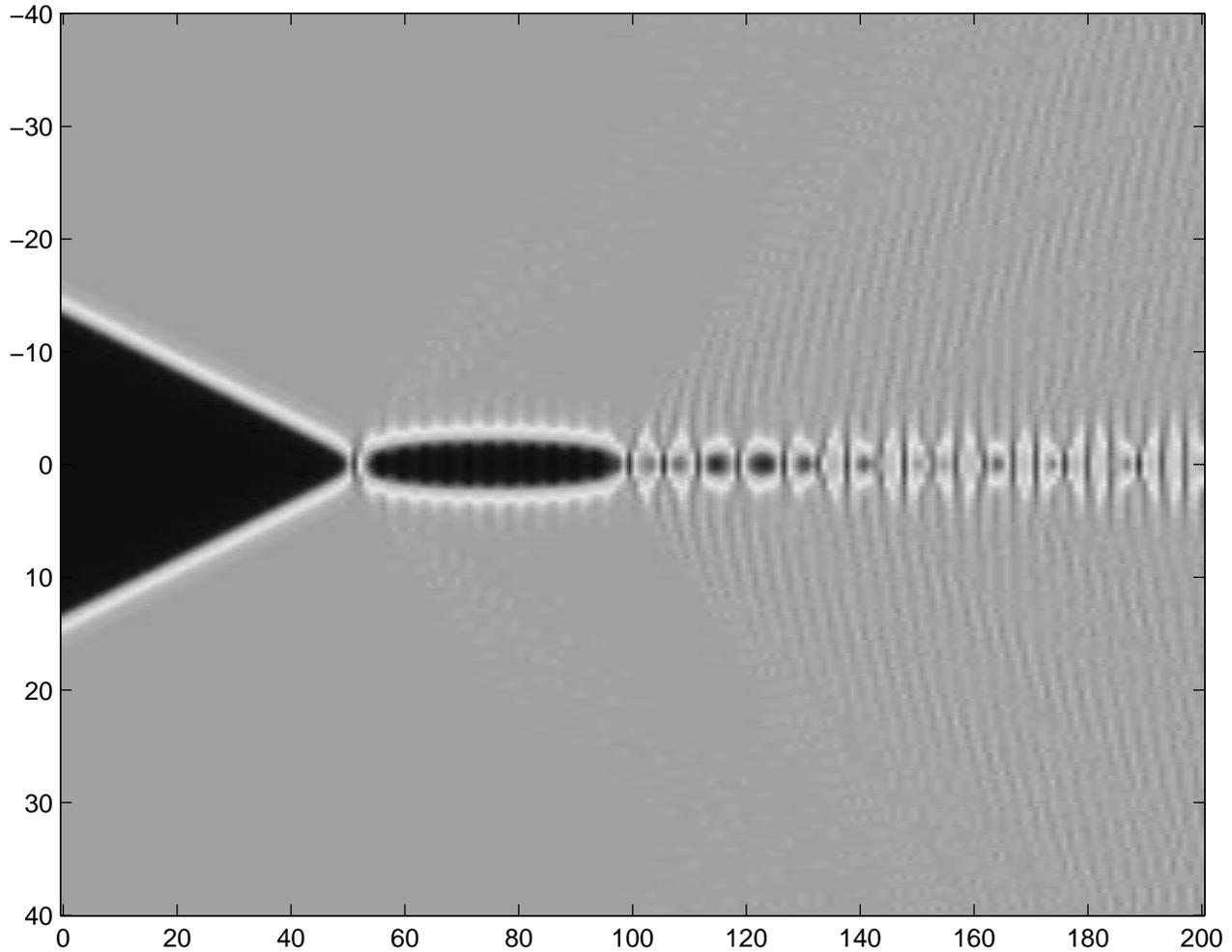}
\caption{Bion formation for model 3 with h = 0.05,
  kink velocity = 0.255, domain size = 80, kink separation = 28. Note
  the radiation (phonons) emanating from the bion.}
%\end{center}
\label{afig12}
\end{figure}

\begin{figure}
%\begin{center}
\includegraphics{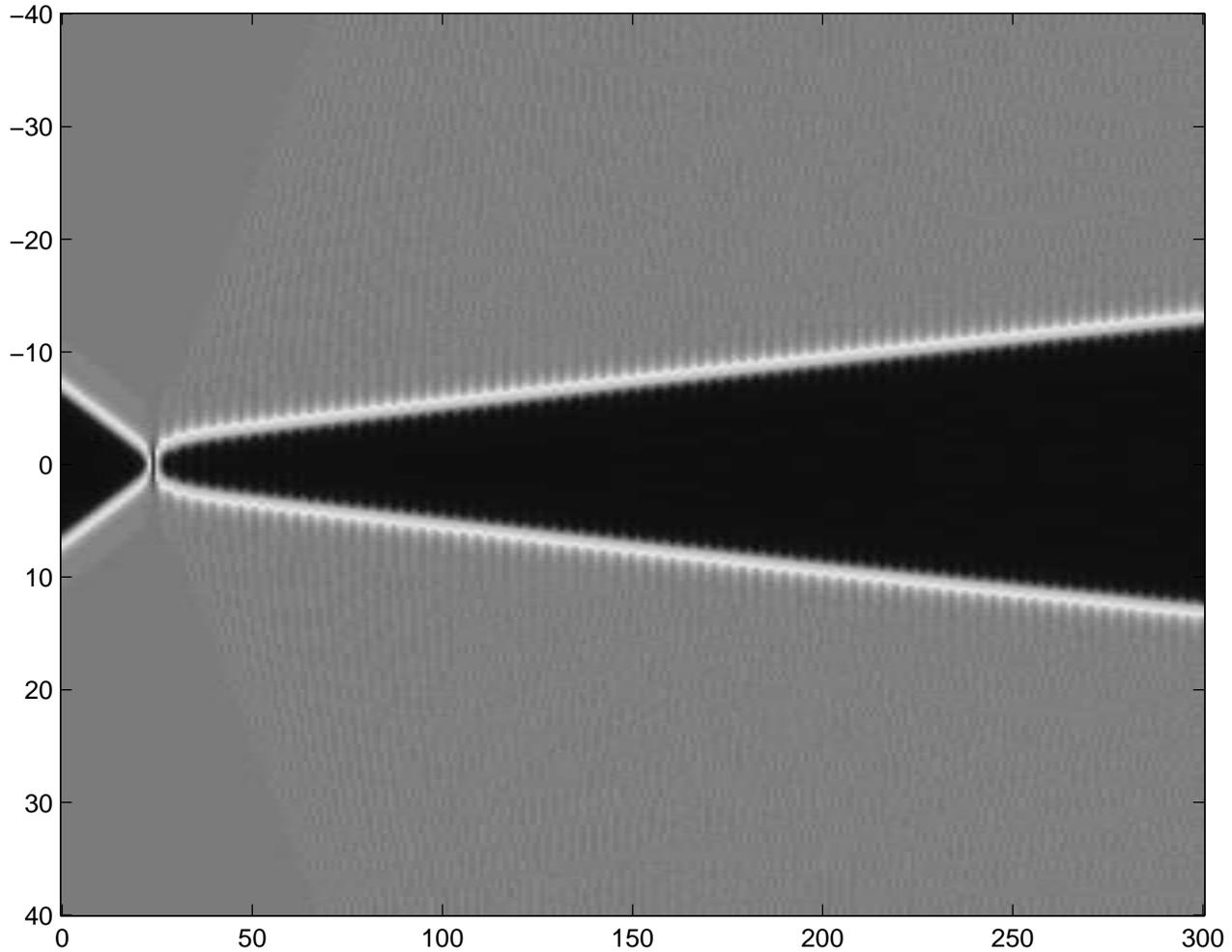}
\caption{One bounce for model 2 with h = 0.1,
  kink velocity = 0.255, domain size = 80, kink separation = 14.}
%\end{center}
\label{afig13}
\end{figure}

\begin{figure}
%\begin{center}
\includegraphics{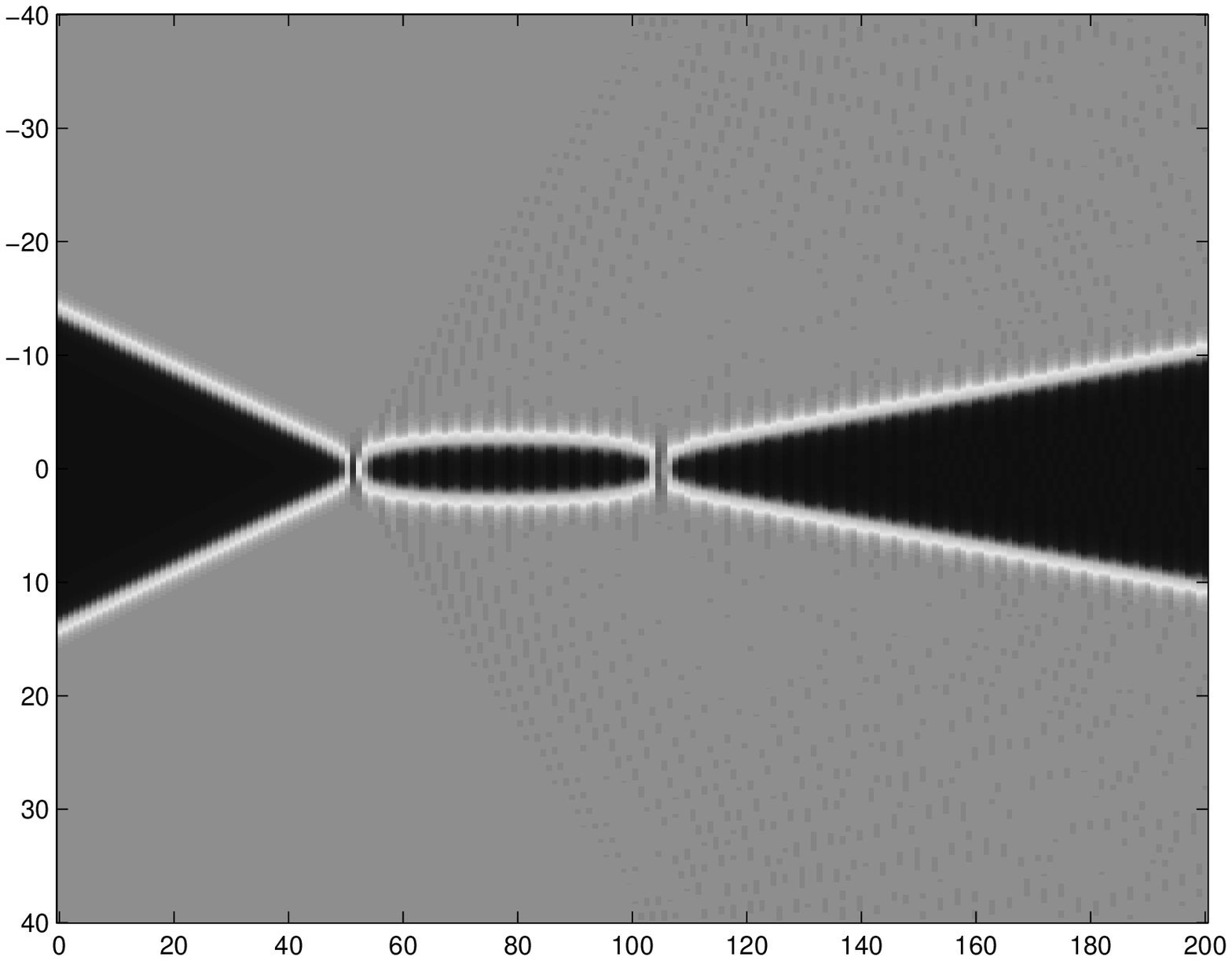}
\caption{Two bounces for model 4 with h = 0.05,
  kink velocity = 0.255, domain size = 80, kink separation = 28.}
%\end{center}
\label{afig14}
\end{figure}

\begin{figure}
%\begin{center}
\includegraphics{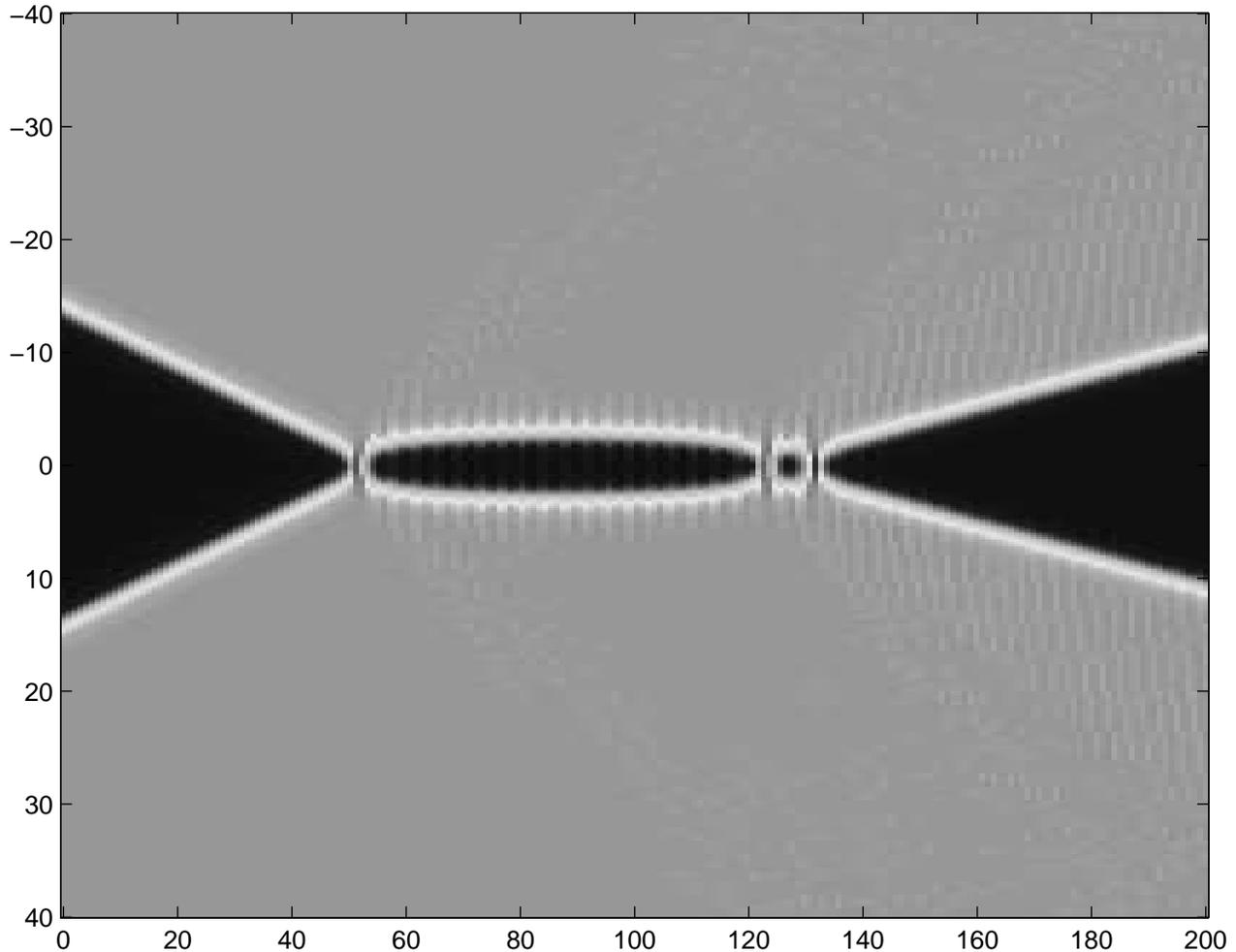}
\caption{Three bounces for model 1 with h = 0.15,
  kink velocity = 0.255, domain size = 80, kink separation = 28.}
%\end{center}
\label{afig15}
\end{figure}

\begin{figure}
%\begin{center}
\includegraphics{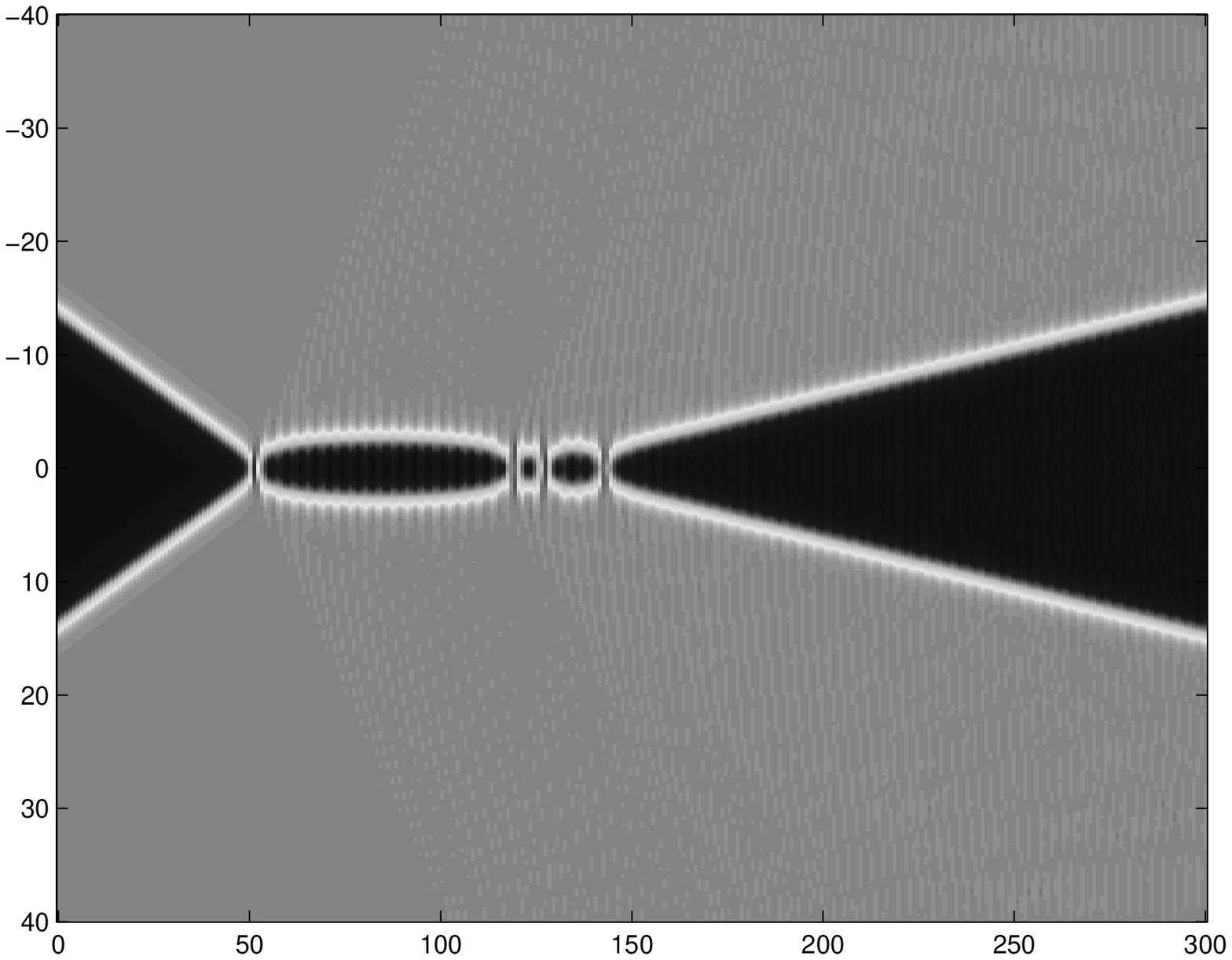}
\caption{Four bounces for model 5 with h = 0.05,
  kink velocity = 0.255, domain size = 80, kink separation = 28.}
%\end{center}
\label{afig16}
\end{figure}

%-------------------------------------------------------------
We have carried out a comparative study of kink collisions under
different discretizations. Our results have been obtained for
different domain sizes (i.e., lattice sizes)
and with different initial separations detailed in Table \ref{tab1}.
As illustrated above, all of our models share the same
continuum limit. We have compared the scattering properties for
four different (dimensionless) velocities $ v= 0.21, 0.225, 0.24$ and $ 0.255$,
respectively presented in Tables \ref{tab2}-\ref{tab5}. We
chose these velocities motivated by their (continuum limit)
phenomenology in the detailed examination
of \cite{anninos}.
In each of the Tables \ref{tab2}-\ref{tab5}, the collision results are shown
with  an increment of $0.025$ in the lattice spacing $h$ for each of the
different selected initial separations and domain sizes.
Some of the standard collision outcomes are highlighted
for $v=0.255$ in Figs.
\ref{afig12}-\ref{afig16}. The most typical cases are those
of figures \ref{afig12}-\ref{afig13}; the former shows a bion formation
(i.e., the kink and the antikink merge, forming an oscillatory,
so-called bion state, and never separate thereafter),
a behavior typical for sufficiently small speeds. The latter
illustrates what is characterized as a ``one-bounce'' separation, a behavior
typical for sufficiently large initial speeds.
However, the delicate structure of collisions for an intermediate
range of speeds may lead to additional fine structure including
multiple bounces before the eventual separation of the two kinks,
as illustrated in Figs. \ref{afig14}-\ref{afig16}.

\begin{table}
\caption{Domain Sizes and Initial Kink Separations (in units of
  lattice constant) Examined \label{tab1}}
\begin{center}
\begin{tabular}{|c|c|c|c|}
\hline
%domain size&separation\\ \hline
domain size&80&80&160\\ \hline
separation&14&28&28\\ \hline
\end{tabular}
\end{center}
\end{table}

{\small
\begin{table}
\caption{Outcome of Kink-Antikink Collisions for $v=0.21$. \label{tab2}}
\begin{center}
\begin{tabular}{|l|c|c|c|c|c|c|c|c|c|c|c|c|c|c|c|}
\hline \multicolumn{16}{|c|}
    {\rule[-3mm]{0mm}{8mm}Results for velocity 0.21} \\ \hline
\multicolumn{1}{|c|}{$h$} &\multicolumn{3}{|c|}{Campbell \it et.al.}
&\multicolumn{3}{|c|}{Speight} &\multicolumn{3}{|c|}{CKMS}
&\multicolumn{3}{|c|}{K1} &\multicolumn{3}{|c|}{K2}\\ \hline
     &80/14&80/28&160/28&80/14&80/28&160/28&80/14&80/28&160/28&80/14&80/28&160/28&80/14&80/28&160/28\\ \hline
0.025&bion&bion&bion&bion&bion&bion&bion&bion&bion&bion&bion&bion&bion&bion&bion\\
\hline 0.05
&bion&bion&bion&bion&bion&bion&bion&bion&bion&bion&bion&bion&bion&bion&bion\\
\hline
0.075&bion&bion&bion&bion&bion&bion&bion&bion&bion&bion&bion&bion&bion&bion&bion\\
\hline 0.1
&bion&bion&bion&bion&bion&bion&bion&bion&bion&bion&bion&bion&bion&bion&bion\\
\hline 0.125&bion&bion&bion&4
&bion&bion&bion&bion&bion&bion&bion&bion&2   &2   &2   \\ \hline
0.15 &bion&bion&bion&4   &4
&bion&bion&bion&bion&bion&bion&bion&4   &bion&bion\\ \hline
0.175&bion&bion&bion&bion&bion&3   &bion&bion&bion&2   &2   &2
&bion&bion&bion\\ \hline 0.2  &bion&bion&bion&bion&bion&2
&bion&bion&bion&bion&bion&bion&bion&2   &2   \\ \hline
0.225&bion&bion&bion&2   &bion&bion&bion&3   &3
&bion&bion&bion&1   &1   &1   \\ \hline 0.25 &3   &bion&bion&1
&1   &1   &bion&bion&bion&2   &2   &2   &1   &1   &1   \\ \hline
0.275&bion&bion&bion&1   &1   &1   &2   &1   &1
&bion&bion&bion&1   &1   &1   \\ \hline 0.3  &2   &bion&bion&1
&1   &1   &bion&1   &1   &bion&2   &2   &1   &1   &1   \\ \hline
0.325&2   &2   &2   &1   &1   &1   &2   &1   &1   &2   &1   &1
&1   &1   &1   \\ \hline
\end{tabular}
\end{center}
\end{table}

\begin{table}
\caption{Outcome of Kink-Antikink Collisions for $v=0.225$. \label{tab3}}
\begin{center}
\begin{tabular}{|l|c|c|c|c|c|c|c|c|c|c|c|c|c|c|c|}
\hline \multicolumn{16}{|c|}
    {\rule[-3mm]{0mm}{8mm}Results for velocity 0.225} \\ \hline
\multicolumn{1}{|c|}{$h$} &\multicolumn{3}{|c|}{Campbell \it et.al.}
&\multicolumn{3}{|c|}{Speight} &\multicolumn{3}{|c|}{CKMS}
&\multicolumn{3}{|c|}{K1} &\multicolumn{3}{|c|}{K2}\\ \hline
&80/14&80/28&160/28&80/14&80/28&160/28&80/14&80/28&160/28&80/14&80/28&160/28&80/14&80/28&160/28\\
\hline 0.025&2   &2   &2   &2   &2   &2   &2   &2   &2   &2   &2
&2   &2   &2   &2   \\ \hline 0.05 &2   &2   &2   &2   &2   &2
&2   &2   &2   &2   &2   &2   &2   &2   &2   \\ \hline 0.075&2
&2   &2   &3   &bion&bion&2   &2   &2   &2   &2   &2
&bion&bion&bion\\ \hline 0.1  &2   &2   &2
&bion&bion&bion&bion&bion&bion&bion&bion&bion&bion&bion&bion\\
\hline 0.125&2   &bion&bion&2   &2   &2   &4
&bion&bion&bion&bion&bion&bion&bion&bion\\ \hline 0.15
&bion&bion&bion&bion&2   &2
&bion&bion&bion&bion&bion&bion&bion&2   &2   \\ \hline
0.175&bion&bion&bion&bion&bion&bion&bion&bion&bion&bion&2   &2
&bion&bion&bion\\ \hline 0.2  &bion&bion&bion&2   &bion&1
&bion&bion&bion&bion&bion&bion&1   &1   &1   \\ \hline 0.225&3
&bion&bion&1   &1   &1   &1   &bion&bion&bion&2   &2   &1   &1
&1   \\ \hline 0.25 &2   &2   &2   &1   &1   &1   &1
&bion&bion&bion&2   &2   &1   &1   &1   \\ \hline
0.275&bion&bion&bion&1   &1   &1   &1   &1   &1   &1   &1   &1
&1   &1   &1   \\ \hline 0.3  &2   &bion&bion&1   &1   &1   &1
&1   &1   &1   &1   &1   &1   &1   &1   \\ \hline
0.325&bion&bion&bion&1   &1   &1   &1   &1   &1   &1   &1   &1
&1   &1   &1   \\ \hline
\end{tabular}
\end{center}
\end{table}

\begin{table}
\caption{Outcome of Kink-Antikink Collisions for $v=0.24$. \label{tab4}}
\begin{center}

\begin{tabular}{|l|c|c|c|c|c|c|c|c|c|c|c|c|c|c|c|}
\hline
\multicolumn{16}{|c|}
    {\rule[-3mm]{0mm}{8mm}Results for velocity 0.24} \\ \hline
\multicolumn{1}{|c|}{$h$}
&\multicolumn{3}{|c|}{Campbell \it et.al.}

&\multicolumn{3}{|c|}{Speight}

&\multicolumn{3}{|c|}{CKMS}

&\multicolumn{3}{|c|}{K1}

&\multicolumn{3}{|c|}{K2}\\ \hline

&80/14&80/28&160/28&80/14&80/28&160/28&80/14&80/28&160/28&80/14&80/28&160/28&80/14&80/28&160/28\\ \hline

0.025&bion&bion&bion&bion&bion&bion&bion&bion&bion&bion&bion&bion&bion&bion&bion\\ \hline

0.05 &bion&bion&bion&bion&bion&bion&bion&bion&bion&bion&bion&bion&bion&bion&bion\\ \hline

0.075&bion&bion&bion&2   &2   &2   &bion&3   &3   &2   &bion&bion&bion&bion&bion\\ \hline

0.1  &bion&bion&bion&3   &2   &2   &bion&2   &2   &bion&2   &2   &bion&2   &2   \\ \hline

0.125&bion&bion&bion&bion&bion&bion&bion&bion&bion&3   &3   &3   &2   &bion&bion\\ \hline

0.15 &bion&2   &2   &2   &1   &1   &bion&bion&bion&bion&3   &3   &1   &1   &1   \\ \hline

0.175&bion&bion&bion&1   &1   &1   &bion&bion&bion&bion&bion&bion&1   &1   &1   \\ \hline

0.2  &2   &2   &2   &1   &1   &1   &2   &3   &3   &1   &2   &2   &1   &1   &1   \\ \hline

0.225&2   &bion&bion&1   &1   &1   &bion&3   &3   &1   &1   &1   &1   &1   &1   \\ \hline

0.25 &bion&2   &2   &1   &1   &1   &bion&bion&bion&1   &1   &1   &1   &1   &1   \\ \hline

0.275&bion&2   &2   &1   &1   &1   &2   &1   &1   &1   &1   &1   &1   &1   &1   \\ \hline

0.3  &bion&bion&bion&1   &1   &1   &1   &1   &1   &1   &1   &1   &1   &1   &1   \\ \hline

0.325&2   &2   &2   &1   &1   &1   &1   &1   &1   &1   &1   &1   &1   &1   &1   \\ \hline

\end{tabular}

\end{center}
\end{table}

\begin{table}
\caption{Outcome of Kink-Antikink Collisions for $v=0.255$. \label{tab5}}
\begin{center}
\begin{tabular}{|l|c|c|c|c|c|c|c|c|c|c|c|c|c|c|c|}
\hline
\multicolumn{16}{|c|}
    {\rule[-3mm]{0mm}{8mm}Results for velocity 0.255} \\ \hline
\multicolumn{1}{|c|}{$h$}

&\multicolumn{3}{|c|}{Campbell \it et.al. }

&\multicolumn{3}{|c|}{Speight}

&\multicolumn{3}{|c|}{CKMS}

&\multicolumn{3}{|c|}{K1}

&\multicolumn{3}{|c|}{K2}\\ \hline

&80/14&80/28&160/28&80/14&80/28&160/28&80/14&80/28&160/28&80/14&80/28&160/28&80/14&80/28&160/28\\ \hline

0.0125&bion&bion&bion&bion&bion&bion&bion&bion&bion&3   &bion&bion&bion&bion&bion\\ \hline

0.025 &bion&bion&bion&2   &bion&2   &bion&bion&bion&bion&bion&bion&2   &2   &2   \\ \hline

0.05  &2   &bion&bion&bion&2   &2   &bion&bion&bion&2   &2   &2   &2   &4   &4   \\ \hline

0.075 &bion&bion&bion&1   &bion&2   &bion&2   &2   &bion&2   &1   &1   &1   &1   \\ \hline

0.1   &2   &bion&bion&1   &1   &1   &1   &bion&bion&1   &1   &1   &1   &1   &1   \\ \hline

0.125 &bion&bion&bion&1   &1   &1   &1   &bion&bion&1   &1   &1   &1   &1   &1   \\ \hline

0.15  &1   &3   &3   &1   &1   &1   &1   &2   &2   &1   &1   &1   &1   &1   &1   \\ \hline

0.175 &1   &bion&4   &1   &1   &1   &1   &2   &2   &1   &1   &1   &1   &1   &1   \\ \hline

0.2   &1   &1   &1   &1   &1   &1   &1   &1   &1   &1   &1   &1   &1   &1   &1   \\ \hline

0.225 &1   &1   &1   &1   &1   &1   &1   &1   &1   &1   &1   &1   &1   &1   &1   \\ \hline

0.25  &1   &1   &1   &1   &1   &1   &1   &1   &1   &1   &1   &1   &1   &1   &1   \\ \hline

0.275 &1   &1   &1   &1   &1   &1   &1   &1   &1   &1   &1   &1   &1   &1   &1   \\ \hline

0.3   &1   &1   &1   &1   &1   &1   &1   &1   &1   &1   &1   &1   &1   &1   &1   \\ \hline

0.325 &1   &1   &1   &1   &1   &1   &1   &1   &1   &1   &1   &1   &1   &1   &1   \\ \hline

\end{tabular}

\end{center}
\end{table}
}

The general trend as displayed in the table \ref{tab2}
is that for velocity $
v= 0.21$, when the kinks collide, they
form a bion state. In the bound state the kink
and the antikink
are trapped by their mutual attraction.
%These trapped states can
%be formed only if the kinks have time to lose sufficient energy in
%a collision.
In the table we see that, for small lattice spacings,
the behavior of the different models is similar (as is expected,
given the common continuum limit); on the other hand,
the dynamics starts to diversify between discretizations, as the spacing
is increased. Remarkably so, for larger values of $h$, we observe
that the collisions are more elastic and, in fact, typically result
in a single bounce for sufficiently large $h$.
%We expect the bion state at low initial velocity and
%inelastic reflection at high relative velocity.
For velocity $ v=
0.225$ (Table \ref{tab3}) the kinks are in a two-bounce window for
small lattice
spacing $h$ but the change in outcome with increasing $h$ is
rather drastic (especially since the two-bounce is a rather
fine-tuned collision outcome, where the internal modes control
the resonant transfer of energy from and back to its original
kinetic form \cite{anninos,Campbell,goodman}).
%Initially the kinks bounce once and they might not have
%enough kinetic energy to escape from the potential well, after the
%second bounce they regain energy and separate.
We see a similar trend for the case with velocity $v=0.24$,
whereby the kinks
form a bion state in the continuum limit but for increasing $h$ we
observe multiple bounces. For higher $h$, the kinks in all five
models collide quasi-elastically i.e., with a single bounce.
For a kink velocity of $v = 0.255$
(which is close to the critical velocity, above which
only single bounce phenomena occur in the continuum)  we see an
additional quite interesting feature. In this case,
the outcomes for different models are so sensitive
that they may not even converge for very small values of
$h$.

%The figures given in the appendix are for kinks colliding with
%initial velocity $v = 0.255$, for different separation and for
%different models. We observe five different outcomes; bion state
%which is a final state containing a spatially localized time
%oscillatory motion or inelastic reflections forming one bounce,
%two bounces, three bounces and four bounces depending on the
%parameters.

Our results overall indicate that the elasticity of the collisions
depends strongly on the lattice spacing as well as on the details
of the particular discretization. The collisions appear to be more
elastic for larger values of $h$, a feature which seems to be
counter-intuitive given that discreteness in this type of models
is perceived as a source of dissipation of kinetic energy
\cite{peyrard1}. On the other hand, discreteness leads to the
excitation of additional internal modes (see Figs. \ref{fig2},
\ref{fig3} and \ref{fig5}, in particular) and hence, potentially,
to more exotic dynamical outcomes of the collisions. Furthermore,
as $h$ increases the width of the phonon band decreases, hence
potentially limiting the range of resonant modes and therefore
the amount of radiated energy (see also the relevant discussion
below). This particular feature (apparent collision elasticity
increase as a function of $h$) would be certainly wortwhile of a
separate and detailed theoretical investigation. From the general
trends of our results, we also observe that the most inelastic
collisions occur for model 1, as might be expected by the presence
of the PN barrier in that model. Finally, one more note of caution
worth making here concerns the disparity between the different
model results even for small $h$. It is clear that such phenomena
as the outcome of collisions depend strongly and sensitively on a
variety of factors (including e.g., the internal mode excitations,
the location of collision, etc.) to an extent that one should not
expect identical collision outcomes among these models even rather
close to the continuum limit (which the models share). This is
also a result that partially defies the conventional wisdom that
would suggest that different collisional outcomes across
discretizations might be expected only when the length scale of
discreteness (the lattice spacing) becomes comparable to the size
of the kink.

In what follows, we briefly analyze one of the sources
of the above mentioned sensitivity of the collision outcome,
namely the original distance between the kink-antikink pair.
We also quantify in a characteristic, in our view, way the
increase in collision elasticity through the dependence
of the critical speed separating bion formation from
one-bounce collisions as a function of the lattice spacing $h$.

%--------------------------------------------------------------------------
\subsection{ Initial distance between colliding kinks}\label{sec:InitDisst}
%--------------------------------------------------------------------------

\begin{figure}
\includegraphics{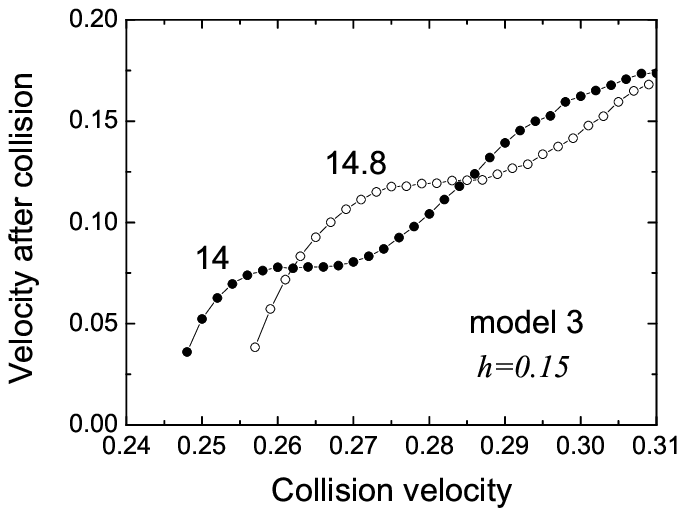}
\caption{Kink velocity after collision as a function of collision
velocity in model 3 at $h=0.15$. Dots show the results for initial
distance between kinks of $14$, while open circles for the initial
distance of $14.8$.} \label{fig8}
\end{figure}

In Fig. \ref{fig8} we show the kink velocity after the collision as a
function of initial collision velocity in model 3 at $h=0.15$. Dots show
the results for initial distance between kinks of $14$, while open
circles for the initial distance of $14.8$.

One can notice a strong sensitivity of the collision outcome to the
initial separation distance. This is because of the internal mode
being excited when boosting the kinks. Changing the initial distance,
we change the phase of the internal mode at the collision point,
which critically, in turn, affects the result of the collision.

%The internal modes can push the kinks or slow them down, depending
%on the phase.

For different models, the sensitivity of the collision outcome to
the initial kink separation correlates with the amplitude of the
kink's internal mode excited at boosting (see Fig. \ref{fig6}).
Thus, the sensitivity is highest for model 3 and lowest for model
5.

The sensitivity also decreases rapidly with decrease in $h$ and the
reason is, essentially, the same: the amplitude of the excited kink's internal
mode decreases as $h^2$.

%--------------------------------------------------------------------------
\subsection{ Threshold velocity $v_c$ as a function of $h$}\label{sec:vch}
%--------------------------------------------------------------------------

\begin{figure}
\includegraphics{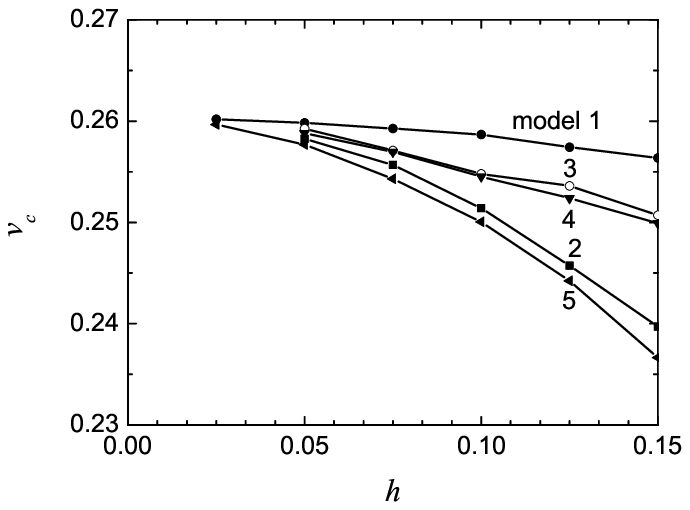}
\caption{Critical velocity $v_c$ as a function of $h$ for the five
  models. In all
models, $v_c$ decreases with increase in $h$ implying that for
larger $h$ collisions are more elastic. The classical discretization
(model 1) shows the weakest dependence of $v_c$ on $h$, while in
the models 2 and 5 this dependence is strongest.} \label{fig10}
\end{figure}

It is well known (see, e.g., \cite{Campbell}) that there exists a
threshold velocity $v_c$ such that collision of kinks with $v>v_c$
leads to separation after the first collision while for collisions
with $v<v_c$,  the first collision cannot lead to separation. In
the latter case, the reflection windows discussed in
\cite{Campbell} can be observed amidst regions of bion formation.

In Fig. \ref{fig10} we show for the five models how $v_c$ changes
with $h$. These results were obtained from computations similar to
the ones presented in Fig. \ref{fig8}; $v_c$ was estimated from the
fit ${\rm constant}\times(v^2-v^2_c)^{1/2}$ suggested in
\cite{Campbell} and thus the effect of the initial separation was
averaged out to some extent.

In all models $v_c$ decreases with increase in $h$ implying that
for larger $h$ the collisions are more elastic. The standard
discretization (model 1) shows the weakest dependence of $v_c$ on
$h$, while in models 2 and 5 this dependence is strongest.

%In the classical work \cite{Campbell} they studied model 1 with
%$h=0.01$ and found $v_c=0.2598$, which is in a good agreement with
%our results.

\begin{figure}
\includegraphics{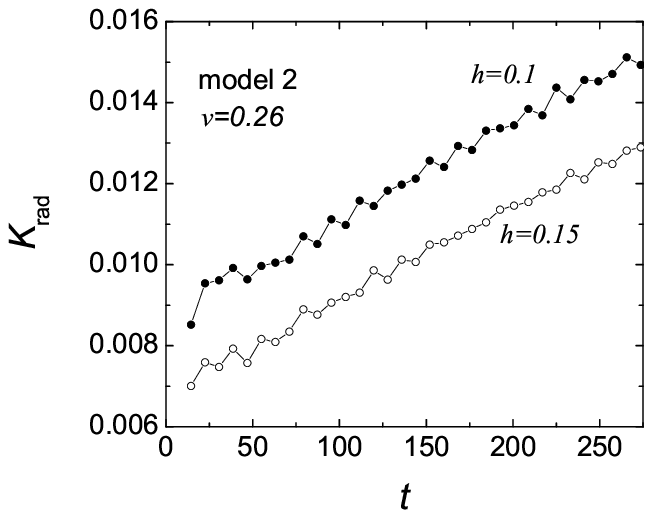}
\caption{Kinetic energy of radiation as a function of the time
after a collision in model 2. Dots show the results for the lattice
spacing $h=0.1$ and open circles for $h=0.15$. The collision velocity
is $v=0.26$. Radiated energy growth with time is due to the
coupling of the kink's internal modes with the phonon band.
Extrapolation of the data to $t=0$ suggests that in the case of
$h=0.1$ collision results in the burst of (dimensionless) kinetic energy of
$8.7\times 10^{-3}$, while a smaller burst of radiation of
$6.8\times 10^{-3}$ takes place in a lattice with higher
discreteness of $h=0.15$.} \label{fig11}
\end{figure}

Since the PN barrier is very small at $h\sim 0.1$, the observed
effect can hardly be explained through the influence of the PN
barrier. As one possible explanation of the dependence of $v_c$ on
$h$, we discuss the burst of radiation emitted during collision.
Corresponding numerical results are presented in Fig. \ref{fig11}
for kinks colliding with $v=0.26$ in the Speight lattice (model 2)
with two lattice spacings, $h=0.1$ and $h=0.15$. This is the
highest velocity we use in our simulations. We show the kinetic
energy of radiation, (in order to exclude the kinetic energy of
the moving kinks, an area of width equal to 4 around each kink was
not included in the computation of the kinetic energy) as a function
of the time after collision. Dots show the results for $h=0.1$ and
open circles for $h=0.15$. The amount of radiated energy grows
with time due to the emission from the kink's internal modes
excited at the collision.

Extrapolation of the data presented in Fig. \ref{fig11} to $t=0$
suggests that, in the case of $h=0.1$, the collision results in the
burst of kinetic energy (in dimensionless units) of $8.7\times
10^{-3}$, while a smaller burst of radiation of $6.8\times
10^{-3}$ takes place in the lattice with higher discreteness of
$h=0.15$. The fact that the burst of radiation is smaller in the
lattice with higher discreteness can be related to the phonon
spectrum width, which decreases with $h$ as $1/h$ for small $h$,
for all five models. The narrower the phonon band, the smaller the
amount of energy that can be radiated and the more elastic the
collision.

%--------------------------------------------------------------------------
\section{Conclusions and Future Challenges} \label{sec:Conclusions}
%--------------------------------------------------------------------------

In the present work, we have analyzed the properties of a number
of recently proposed discretizations of the continuum $\phi^4$
field theory in the vicinity of (and further away from) the continuum
limit. The relevant
analysis consisted of the examination of the static properties of
the models, concerning their fundamental nonlinear wave solutions,
namely the kinks (and anti-kinks). For these types of solutions,
we have examined how to obtain them, in what ways they differ from
their continuum siblings, as well as the spectral properties of
the linearization around such solutions. In particular, we have
computed both the phonon (continuous) spectrum, as well as
discussed the internal (or shape) modes present in the models.

On the other hand, we have also examined dynamic properties of the
kinks by studying their collision and comparing/contrasting their
outcomes across the different discretizations. In that regard, we
have observed a variety of interesting results. In particular, we
have seen that the models only align with their continuum limit
(especially as regards sensitive collision phenomenology) {\it
extremely close} to the continuum limit [(i.e., for spacings of
O$(10^{-2})$)]. In fact, in some cases (e.g. for the CKMS model and
$v=0.255$), the results are not independent of factors such as
lattice spacing and kink-antikink separation even for the smallest
$h$ used herein ($h=0.0125$). This is rather remarkable given that
the scale of the kinks themselves is considerably wider, hence one
would not expect this result on the basis of length-scale
competition. However, we have argued that this should be
attributed to the (initial-boost induced) excitation of the
internal modes of the kink whose coupling to the continuous
spectrum sensitively affects the collision outcome, as has been
substantiated previously
\cite{anninos,Campbell,Campbell1,goodman}. This should operate as
a significant note of caution to researchers conducting numerical
experiments with these models in an attempt to describe their
continuum limits.

Furthermore, we have seen that the elasticity of collisions varies
not only from model to model, but even with increasing spacing of
the lattice. In particular, we have illustrated through our
numerical observations that the most inelastic collisions take
place in the model that does have a Peierls-Nabarro barrier, while
PNb-free models feature more elastic collisions. Moreover, there
is a very interesting (and worthwhile to investigate further,
possibly theoretically as well) dependence of the critical speed
for single-bounce collisions on the spacing $h$. In particular,
$v_c$ rapidly decreases as a function of $h$, rendering
coarser collisions more elastic. This should also be attributed to
the spectral properties of the models and the decreasing width of
the phonon band for increasing $h$, which activates fewer
couplings of internal mode frequency harmonics with the continuous
spectrum and hence leads to weaker ``dissipation'' and consequently to
more elastic collisions.

While the static properties of the kinks have been obtained to a
large extent explicitly from the underlying discretized first
integral formalisms, kink collisions are naturally much harder to
analyze theoretically for the presented discrete models. However,
some of the relevant features such as the dependence of $v_c$ on
$h$ may be, to a certain degree, tractable (see e.g.
\cite{goodman} and references therein). Hence, it would be
particularly interesting to seek a deeper understanding
of the features numerically observed herein and how these can be
associated with the nature of the underlying discretized
nonlinearity. Such studies are currently in progress.

\section*{Acknowledgements} I. Roy gratefully acknowledges
the hospitality of the Center for Nonlinear Studies at Los
Alamos National Laboratory. PGK gratefully acknowledges support
from NSF-DMS-0204585, NSF-DMS-0505663 and NSF-CAREER. Work at Los
Alamos was performed under the auspices of the U.S. Department of
Energy.


\begin{thebibliography}{99}


\bibitem{konotop} V. V. Konotop and V. A. Brazhnyi,
Mod. Phys. Lett. B {\bf 18} 627, (2004); P.G. Kevrekidis and D. J.
Frantzeskakis, Mod. Phys. Lett. B {\bf 18}, 173 (2004).


\bibitem{nlo}  D. N. Christodoulides, F. Lederer, and Y. Silberberg,
Nature \textbf{424}, 817 (2003); Yu. S. Kivshar and G. P. Agrawal,
\textit{Optical Solitons: From Fibers to Photonic Crystals},
Academic Press (San Diego, 2003).

\bibitem{peyrard} M. Peyrard, Nonlinearity {\bf 17}, R1 (2004).

\bibitem{kivshar1} O.~M.~Braun and Y.~S.~Kivshar, {\it The
Frenkel-Kontorova Model: Concepts, Methods, and Applications}
(Springer, Berlin, 2004).


\bibitem{belova} T. I. Belova and A. E. Kudryavtsev,
Phys. Usp. {\bf 40}, 359 (1997).


\bibitem{anninos} P. Anninos, S. Oliveira, and R.A. Matzner,
Phys. Rev. D {\bf 44}, 1147 (1991).

\bibitem{Campbell} D. K. Campbell, J. F. Schonfeld, and C. A.
Wingate, Physica D {\bf 9}, 1 (1983).

\bibitem{Campbell1} D.K. Campbell and M. Peyrard,
Physica D {\bf 18}, 47 (1986); {\it ibid.} {\bf 19}, 165 (1986).

\bibitem{goodman} R.H. Goodman and R. Haberman,
SIAM J. Appl. Dyn. Sys. {\bf 4}, 1195 (2005).

\bibitem{sulem}  C. Sulem and P.L. Sulem,
\newblock {\it The Nonlinear Schr{\"o}dinger Equation},
(Springer-Verlag, New York, 1999).


\bibitem{SpeightKleinGordon} J. M. Speight, Nonlinearity {\bf 12}, 1373 (1999).

\bibitem{Speight} J. M. Speight and R. S. Ward, Nonlinearity {\bf 7}, 475 (1994).

\bibitem{SpeightPhi4} J. M. Speight, Nonlinearity {\bf 10}, 1615
(1997).

\bibitem{Bogom} E. B. Bogomol'nyi, J. Nucl. Phys. {\bf 24}, 449
(1976).

\bibitem{KevrekidisPhysD} P. G. Kevrekidis, Physica D {\bf 183}, 68 (2003).

\bibitem{Saxena} F. Cooper, A. Khare, B. Mihaila, and A. Saxena,
Phys. Rev. E {\bf 72}, 36605 (2005).

\bibitem{JPhysA} S. V. Dmitriev, P. G. Kevrekidis, and N.
Yoshikawa, J. Phys. A {\bf 38}, 7617 (2005).

\bibitem{Barashenkov} I. V. Barashenkov, O. F. Oxtoby, and
D. E. Pelinovsky,  Phys. Rev. E {\bf 72}, 035602(R) (2005);
 O. F. Oxtoby, D. E. Pelinovsky, and I. V. Barashenkov,
Nonlinearity {\bf 19}, 217 (2006).

\bibitem{dk}  S.V. Dmitriev, P.G. Kevrekidis, A.A. Sukhorukov,
N. Yoshikawa, and S. Takeno, Phys. Lett. A {\bf 356}, 324 (2006);
P. G. Kevrekidis, S. V. Dmitriev, and A. A. Sukhorukov,
nlin.SI/0603046.

\bibitem{dep}  D.E. Pelinovsky,
nlin.PS/0603022.

\bibitem{krss} A. Khare, K. \O. Rasmussen, M. R. Samuelsen, and A.
Saxena, J. Phys. A: Math. Gen. {\bf 38}, 807 (2005); A. Khare, K. \O.
Rasmussen, M. Salerno, M. R. Samuelsen, and A. Saxena, Phys. Rev. E
{\bf 74}, 016607 (2006).

\bibitem{almeida} A.B. Adib and C.A.S. Almeida,
Phys. Rev. E {\bf 64}, 037701 (2001).

\bibitem{AL} M. J. Ablowitz and J. F. Ladik,
\newblock J. Math. Phys. {\bf 16}, 598 (1975);
M. J. Ablowitz and J. F. Ladik,
\newblock J. Math. Phys. {\bf 17}, 1011 (1976).

\bibitem{DKYF} S. V. Dmitriev, P. G. Kevrekidis, N. Yoshikawa,
and D. J. Frantzeskakis, nlin.PS/0603074.

\bibitem{pla} P. G. Kevrekidis, C.K.R.T. Jones,  and T. Kapitula,
Phys. Lett. A {\bf 269}, 120 (2000).

\bibitem{sugiyama} T. Sugiyama, Prog. Theor. Phys.
{\bf 61}, 1550 (1978).

\bibitem{peyrard1}  M. Peyrard and M.D. Kruskal,
Physica D {\bf 14}, 88 (1984).

\end{thebibliography}
\end{document}